\documentstyle[pre,aps]{revtex} 
\input epsf
\begin{document}
\draft
\twocolumn[\hsize\textwidth\columnwidth\hsize\csname @twocolumnfalse\endcsname

\title{Field-induced Ordering in Critical Antiferromagnets}

\author{S.\ L.\ A.\ de Queiroz, Thereza Paiva, Jorge S.\ de S\'a Martins, 
and Raimundo R. dos Santos}
\address{
Instituto de F\'\i sica, Universidade Federal Fluminense, 
Avenida Litor\^anea s/n, 24210-340 Niter\'oi RJ, Brazil}
\date{\today}
\maketitle
\begin{abstract}
Transfer-matrix scaling methods have been used to study
critical properties of field-induced phase transitions of
two distinct two-dimensional antiferromagnets with discrete-symmetry
order parameters:  triangular-lattice Ising systems (TIAF) and the
square-lattice three-state Potts model (SPAF-3). 
Our main findings are summarised as follows.
For TIAF, we have shown that the critical line leaves the zero-temperature, zero-field fixed point at a finite angle. Our best estimate
of the slope at the origin is $\left(dT_c/dH\right)_{T=H=0} = 4.74 \pm 0.15$.
For SPAF-3 we provided evidence that the zero-field correlation length
diverges as $\xi(T \to 0, H=0) \simeq \exp (a/T^{x})$, with $x=1.08 \pm 0.13$,
through analysis of the critical curve at $H \neq 0$ plus crossover
arguments. For SPAF-3 we have also ascertained that the conformal anomaly
and decay-of-correlations exponent behave as:   
(a) $H=0$: $c=1$, $\eta=1/3$; (b) $H \neq 0$: $c=1/2$, $\eta=1/4$. 
\end{abstract}

\narrowtext
\vskip2pc]

\section{Introduction}
\label{intro}

Frustrated systems with macroscopically degenerate ground states 
still pose intriguing theoretical challenges. 
For pure Ising and $q$-state Potts antiferromagnets (AF's), in particular,
this degeneracy results solely from the interplay between ``dynamics'' 
(i.e., the number of states per lattice site) and geometry (lattice topology).
In this paper we shall deal exclusively with
Ising and 3-state Potts 
AF's, respectively on the triangular and square lattices. 
It is well established that, in the absence of an external magnetic field,
both systems display
a critical ground state (in the sense that spin-spin correlations decay 
algebraically with distance), and are paramagnetic for all temperatures 
$T>0$~\cite{Berker80,Baxter82,Night82,dN82,Steph64}.
In both cases, a uniform field $H$ removes the residual entropy per spin, 
and 
long range order can set in at finite temperatures, below a field-dependent 
phase boundary $T_c(H)$~\cite{Kinzel81,Racz83}. 

For the Ising antiferromagnet on the triangular lattice (TIAF), 
the phase transition at $T_c(H)$ has been determined to be in the 3-state 
ferromagnetic Potts model universality class~\cite{Kinzel81,alex75}; 
subsequent analysis in the context of conformal invariance~\cite{Noh92} led 
to a conformal anomaly (or central charge)
$c=4/5$, which is also consistent with the value
for the 3-state Potts ferromagnet~\cite{bcn}.
Further, it has been found that the critical phase at $T=H=0$
extends into a small region 
$T=0,\ H\leq H_{KT}\simeq 0.27$~\cite{Nienhuis84,Blote91,Blote93};
that is, $H$ is not a {\em relevant} scaling field at $T=0$, as initially
thought~\cite{Kinzel81}.
At $H_{KT}$ the system undergoes a Kosterlitz-Thouless (KT) transition
to a long-range ordered state~\cite{Blote93}. Within the zero-temperature
critical phase, one has continuously-varying critical exponents and, 
accordingly~\cite{cardy87}, the conformal anomaly is $c=1$~\cite{Blote93}.
Much less is known about field effects on the 3-state Potts antiferromagnet 
on the square lattice (SPAF-3), apart from indications that, for $H \neq 0$
the transition at the corresponding $T_c(H)$ belongs to the
two-dimensional Ising model universality class~\cite{Racz83}.

Since the scaling behaviour (i.e., relevance or marginality) of the uniform 
field at $T=0$ influences
the shape of the critical curve near $T=H=0$ in a fundamental way, 
an accurate evaluation of $T_c(H)$ is clearly of interest. 
With this in mind, here we investigate the finite-temperature field-induced 
transition in both the TIAF and the SPAF-3, by means of transfer-matrix 
scaling methods~\cite{Night76,Night90}.
For TIAF we concentrate on the shape of the critical curve close to $T=H=0$,
for reasons to be stated in the corresponding section.
For SPAF-3  we determine the critical curve, as well as the conformal anomaly
and the decay-of-correlations exponent $\eta$ along it.
The layout of the paper is as follows. 
In Sec.\ \ref{calcs} we outline our calculational 
procedure for the free energy and the correlation length, from which we
determine the conformal anomaly and the critical curve, respectively.
Results for the TIAF and SPAF-3 are presented in Sections \ref{TRI}
and \ref{SQ}, respectively.
Sec.\ \ref{conc} summarizes our findings.

\section{Models and Transfer Matrix Scaling}
\label{calcs}

We consider infinitely long strips of width $L$, with periodic boundary 
conditions in both directions.
Ising or Potts spins sit on lattice sites and interact with each other,
as well as with a uniform field, according to the Hamiltonian 
(including the multiplicative factor $-\beta=-1/k_{\rm B}T$),
\begin{equation}
{\cal H} = -K \sum_{\langle i, j\rangle}\ \delta_{\sigma_i \sigma_j}
          + H \sum_i \delta_{\sigma_i0},
\label{Ham}
\end{equation}
where the first sum runs over nearest neighbour sites of either a triangular 
or a square lattice, depending, respectively, on whether $\sigma_i$ is taken 
to be 0 or 1 (Ising), or 0,1, or 2 (3-state Potts); 
a convenient strip geometry for a triangular lattice corresponds to
the usual square strip with additional bonds along a fixed diagonal 
direction.
$K$ is the exchange coupling constant and the field $H$ has been 
taken along the 0-direction.

We shall use units in which, for the triangular Ising {\it ferro}magnet,
$T_c^{-1} = K_c = {1 \over 2}\ln\sqrt{3}$, and for TIAF the upper 
critical field [such that 
$T_c(H \geq H_c) = 0$] is $H_c =6$~\cite{Kinzel81,Noh92}. For SPAF-3,
the corresponding quantities are $T_c^{-1} = \ln(\sqrt{3}+1)$ (ferromagnet), 
$H_c=4$~\cite{Racz83}. 

As usual~\cite{Night76,Night90}, the free energy per spin $f_L(T,H)$
and the correlation length $\xi_L(T,H)$  are given by
\begin{equation}
f_L(T,H)=-{\zeta \over \beta } \ln \lambda_1;\ \ \ \xi_L^{-1}(T,H) = 
-\zeta \ln \left(\lambda_2 / \lambda_1\right)
\label{gibbs}
\end{equation}
where $\lambda_1$ ($\lambda_2$) is the largest (second-largest) eigenvalue 
of the transfer matrix between two successive columns;  the geometric
factor $\zeta =2/\sqrt{3}$ for triangular, and 1 for square lattices. 

We have obtained finite-size estimates of the critical line by
standard phenomenological renormalization (PRG) 
procedures~\cite{Night76,Night90}: for fixed $H$ we consider pairs of strips
of respective widths $L$ and $L^{\prime}$ and solve the implicit
equation
\begin{equation}
L\ \xi_L^{-1}(T^{\ast},H) =  L^{\prime}\ \xi_{L^{\prime}}^{-1}(T^{\ast},H)
\label{prg}
\end{equation}
\noindent for the fixed-point temperature $T^{\ast}$. This approach is rather
safe because the only underlying assumption is that a second-order
phase transition occurs, without any further hypothesis  on its universality
class. As explained  in detail below, we will be particulary concerned
with the proximity of $(T=0,H=0)$, where crossover between different
sorts of critical behaviour is expected. Owing to sublattice symmetries,
$L$ and $L^{\prime}$ must be both
multiples of 3 for TIAF (where we use $L^{\prime} = L-3$, $L = 6, \dots, 18$),
and 2 for SPAF-3 (respectively $L^{\prime} = L-2$, $L=4, \dots, 12$).
At some special points, such as $(T=0,H=0)$, we went up to
$L=14$ for SPAF-3.

For each fixed $H$ the sequence of finite-$L$ estimates of $T^{\ast}$
was extrapolated 
against suitable inverse powers of $L$, so that a set of temperatures
$T_c^{extr} (H)$ was produced, which represents our best approximation to
the true critical curve. We then calculated the free energy and the
correlation length for finite $L$, as in Eq. (\ref{gibbs}), 
at $T_c^{extr} (H)$. From these we produced finite-size estimates of the
conformal anomaly $c$ and the decay-of-correlations exponent $\eta$, 
respectively via
\begin{equation}
L^2\left[f_L(T_c)-f_{\infty}(T_c)\right] = -{\pi c \over 6},
\label{ca}
\end{equation}
and
\begin{equation}
\eta = {L \pi \over \xi(T_c)},\ \ 
\label{etac} 
\end{equation}
as given by conformal invariance~\cite{bcn,cardy84},
where the $H$- dependence of $T_c$ (and thus, of $c$ and $\eta$) 
is implicitly understood.
The sequences of finite-$L$ estimates $c_L$ and
$\eta_L$ were again extrapolated to $L \to \infty$ to give the final values
$c(H)$ and $\eta(H)$.

The extrapolated phase boundaries were usually obtained under the simplifying
assumption of single-power corrections to scaling: for given
$H$ we assumed 
\begin{equation}
T^{\ast}(L,H) - T_c(H) \sim L^{-\psi}\ \  \ .
\label{deltatc}
\end{equation}
\noindent It is expected~\cite{dds,luck} that $\psi = \omega + 1/\nu$, 
where $\omega$ is the leading correction-to-scaling exponent:
\begin{equation}
\xi_L(T_c) = A_0 L \left( 1 + A_1 L^{-\omega} + \ldots \right)
\label{omega}
\end{equation}
and $\nu$ is the correlation-length critical exponent, known to be 5/6 (1)
for 3-state Potts (Ising) ferromagnets  [relevant for TIAF (SPAF-3)
in non-zero field]. However, $\omega$ is not known in advance: though in
several cases\cite{bdn88} numerical evidence has been given in support
of $\omega =2$, this is in principle a non-universal quantity
which depends on the pertinent operator algebra, and probably on lattice
effects as well; {\it e.g.}, for Ising
ferromagnets on triangular and honeycomb lattices, it has been found\cite{unpub}
that $\omega =4$ fits Eq.\ (\ref{omega}) extremely well. Attempts to keep
$\omega$ fixed at 2 in Eq.\ (\ref{deltatc}) for the present case
resulted in fits of widely varying quality.
Thus we took a pragmatic view, and for each $H$ varied $\psi$ 
within reasonable limits (to be spelt out below) until a good fit turned out.
More often than not, the smallest--$L$
term [$L=6(4)$ for TIAF (SPAF-3)] was discarded. Typical uncertainties for
$T_c^{extr}(H)$ were one part in $10^4$, which means that the dominant 
contributions to final spreads in $\eta(H)$ and $c(H)$ are attributable
to the respective extrapolations of $\eta_L$ and $c_L$ to $L \to \infty$ .
From Eq.\ (\ref{omega}), the corrections to $\eta_L$ of
Eq.\ (\ref{etac}) are expected to behave as $L^{-\omega}$ . Surprisingly,
a fixed $\omega=2$ gave reasonably good fits throughout the range of
fields investigated, for both models. For the conformal anomaly
the additional unknown $f_{\infty}(T_c)$ arises;  
assuming corrections to scaling to Eq.~(\ref{ca}) also  with $\omega=2$,
(in this case, such corrections have been shown to work well for 
Potts~\cite{bn82} and Ising~\cite{dQ95} ferromagnets) we performed 
least-squares fits of our data to a parabolic form in $L^{-2}$~\cite{dQ95}.

Further, the field dependence of $c$ and $\eta$ can be analysed within
a finite-size scaling (FSS) theory of crossover effects\cite{rrds81}.
We first assume the existence of two bulk correlation lengths, $\xi^0$ and 
$\xi^T$: the former is only divergent at $T\to 0,\ H=0$;
the latter diverges both at $T\to 0,\ H=0$ (i.e., $\xi^0\sim\xi^T$ in this
case), as well as at $T_c(H)>0$, with different asymptotic forms. 
The two scaling variables are then $L/\xi^0$ and $L/\xi^T$, 
which allows us to cast the finite-size correlation length and free energy 
in the forms 
\begin{equation}
\xi^T_L(T,H)=L\ Q(L/\xi^0,L/\xi^T),
\label{xoverxi}
\end{equation}
and 
\begin{equation}
f_L(T,H)=L^{-d}\ R(L/\xi^0,L/\xi^T),
\label{xoverf}
\end{equation}
respectively, where $Q$ and $R$ are extended scaling functions. 
The crucial difference between the scaling of these and of 
any other quantity ({\em e.g.} susceptibility, specific heat, magnetization) 
is that the leading power in the $L$--dependence is fixed,
instead of a ratio of critical exponents, $x/\nu$, which may 
change from $x_0/\nu_0$ to $x_T/\nu_T$. Indeed, the main $H$-dependence
in the crossover function 
is expected to arise as $[f(H)]^{\epsilon}$, with $\epsilon = x_T/\nu_T -
x_0/\nu_0$~\cite{rrds81}. Thus, low-order corrections to the
asymptotic behaviour must be wiped out; the field-dependent
crossovers in both $\eta$
and $c$ are therefore expected to be quite fast. We shall see that
these predictions are borne out rather well by numerical data for SPAF-3.
For TIAF, technical difficulties (to be described) connected to extrapolation
of the critical boundary translate into a more mixed picture.

\section{Triangular Ising Antiferromagnet}
\label{TRI}
For TIAF in the
range $1 \leq H \leq 5.5$, best fits to Eq. (\ref{deltatc}) were attained with
$\psi \sim 3.5 - 5.5$ (higher values for lower fields). In that region our PRG
estimates extrapolate to values virtually identical to those found in
Ref.\onlinecite{Noh92}. Those authors started from the assumption that,
for all $H \neq 0$, the TIAF is in the same universality class as the 3-state
two-dimensional Potts ferromagnet~\cite{Kinzel81,alex75}, and located the points
$(T,H)$ where $\xi_L(T,H)= 15L/4\pi$, corresponding to $\eta=4/15$ as given
by conformal invariance~\cite{cardy84}. Procedures of this sort were
put forward by Bl\"ote and den Nijs~\cite{bdn88}, and are
expected to be less vulnerable
than PRG to numerical inaccuracies, provided the universality class of
the transition is not in doubt. However, 
near a multicritical point (such as $T=H=0$ here) crossover effects
may also take their toll. Indeed, even though there is no {\em a priori} 
reason to question universality in the present case, convergence of the data of
Ref.\onlinecite{Noh92} deteriorates rapidly close to the origin, to such
an extent that the authors quote no extrapolations for the
critical line for $H \leq 0.5$. In our investigation, we found that for
$ 0.15 \lesssim H \lesssim 0.5$ the PRG curves crossed each other,
thus making the extrapolation procedure unworkable. An example can
be seen near the right edge of Fig.\ \ref{fig:tiafpd} below, which also
shows that closer to the origin the curves again behave monotonically against
$L$.

Before giving details of extrapolation in that region, we recall that
the shape of the low-$T$, low-$H$ phase boundary was discussed
in Ref.\onlinecite{Kinzel81}.
At the time it was believed that $H$ was
a relevant scaling field along $T=0$, from which it was concluded that, since the correlation length diverges as $\exp (a/T)$~\cite{Steph70} at $H=0$,
$T \to 0$, the critical line should approach the origin tangentially to the 
$T$- axis.
On the other hand, PRG with $L=6$ and 9, in the notation of our Eq.~(\ref{prg}),
yielded a finite slope at the origin.
This was interpreted as a deficiency of the
calculational method~\cite{Kinzel81}.
Here we have reexamined the matter by extending PRG to 
$L=18$, extrapolating our data and making contact with the more recent
results~\cite{Blote91,Blote93} which
point to the existence of a Kosterlitz-Thouless phase at $T=0$, $H \neq 0$.

\begin{figure}
\epsfxsize=8,5cm
\begin{center}
\leavevmode
\epsffile{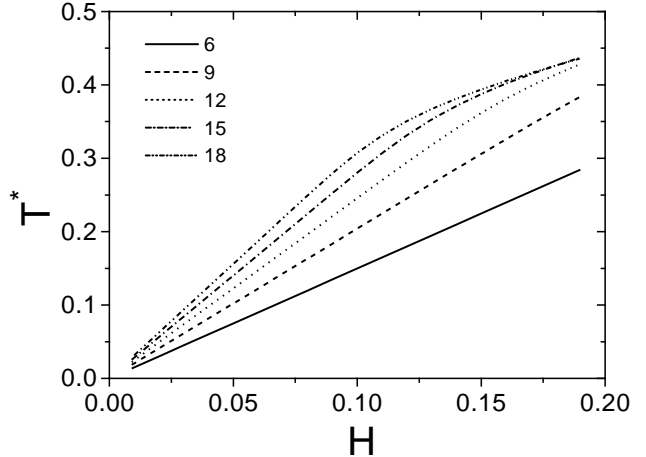}
\caption{PRG estimates of the critical line of TIAF near the origin,
obtained by solving 
Eq.\ (\protect{\ref{prg}}).
$L$ values given on figure ($L^{\prime} =L-3$).}
\label{fig:tiafpd}
\end{center}
\end{figure}

In Figure \ref{fig:tiafpd} the fixed-point solutions of Eq.(\ref{prg}) are
displayed. Though convergence becomes prohibitively slow for $H \lesssim
8 \times 10^{-3}$, there is plenty of leeway to establish that, as $H \to 0$,
all our finite-$L$ curves become straight lines which (to within one part in
$10^4 - 10^5$) cross the origin. The straight sections become shorter with
growing $L$, thus one must be careful before predicting a finite slope
at the origin for the actual phase diagram. In Table \ref{tab:tiaf} we
show the slopes $S_L$ of the straight sections of finite-$L$ curves, as well as
the values $H_{max}(L)$ above which said curves begin to deviate from linear
behaviour. Error bars for $H_{max}(L)$ are somewhat subjective, but 
certainly quite conservative, as can be seen from visual inspection. Should 
$L \to \infty$ extrapolation produce a definitely negative value of $H_{max}$,
one could be sure that the finite slope is a finite-size
artifact. However, we have found $H_{max}(\infty) = 0.01 \pm 0.02$ from a
rather good scaling of our data against $L^{-1}$. 

Bearing in mind that the only ``small'' typical field naturally
arising in the problem, $H_{KT}$, is one order of magnitude larger than this
(thus a strictly positive $H_{max}(\infty)$ of order $10^{-2}$
would have no clear physical origin),
we interpret the above result as signalling that $H_{max}(\infty)$ is
{\em exactly} zero. So, $(i)$ the critical line does
leave the origin at a finite slope (for which our best estimate, $4.74 \pm
0.15$, comes from extrapolation of the $S_L$ against $L^{-2}$);
but $(ii)$ the straight-line part of the critical curve is of zero extent:
one only has $d^2 T_c / dH^2 = 0$ at the origin. This latter quantity 
must be negative for all $H>0$, as no inflection points are expected.     

These conclusions are consistent with the presence of a critical phase
on the $H$- axis near the origin. The exponentially diverging
correlation length~\cite{Steph70} at $H=0, T \to 0$ is roughly in balance
with the (already infinite) $\xi_{KT}$ along the zero-temperature axis. This way
the critical line, where crossover between temperature-- and field--dominated 
behaviour takes place, starts at finite angles with both axes. 

Going back to extrapolation of the critical line for $0 < H \lesssim 0.15$,
we first note that, by construction, our procedure of fixed-$H$
extrapolation automatically yields a straight line with slope $4.74 \pm 0.15$
for all $H \leq H_{max} (18) \simeq 0.09$. From the preceding arguments on the
extent of the straight-line part, and concavity, of the critical curve,
this is an upper limit: $T_c^{real}(H) \leq T_c^{extr}(H)$.
Also, for $0.09 \leq H \lesssim 0.15$ we have not managed to produce good
fits with single powers; instead, we were forced to resort to two-power fits
using $L^{-1}$ and $L^{-2}$. These facts have strong effects on the evaluation 
of $c$ and $\eta$ near $H=0$, which we now turn to discuss.

\begin{figure}
\epsfxsize=8,5cm
\begin{center}
\leavevmode
\epsffile{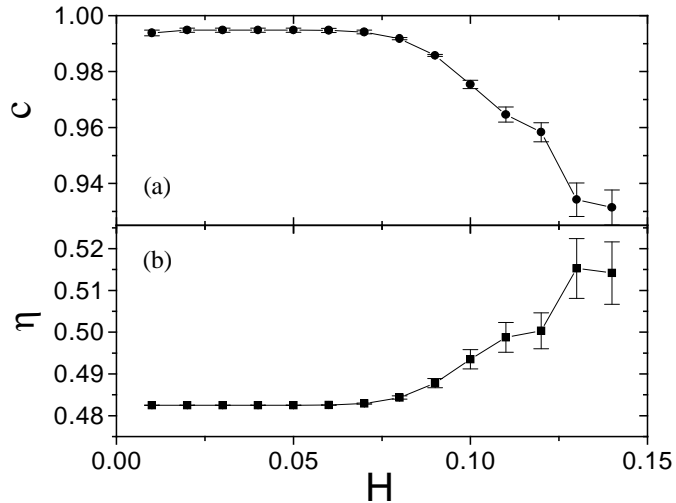}
\caption{Conformal anomaly $c$ (upper curve) and exponent $\eta$ (lower curve)
along the extrapolated critical line of TIAF near the origin. Expected values
are (a) $H=0$: $c=1$, $\eta=1/2$; (b) $H \neq 0$: $c=4/5$, $\eta=4/15$.}
\label{fig:tretac} 
\end{center}
\end{figure}

Recall that at $T=H=0$ one 
has~\cite{Steph70,Blote93}  $c=1$ and $\eta=1/2$, while for $H \neq 0$
the 3-state Potts values $c=4/5$, $\eta= 4/15$ are believed to 
hold~\cite{Noh92}. 
Our results for $c$ and $\eta$ along the extrapolated critical line, near
the origin, are shown in Figure  \ref{fig:tretac}. 
From the discussion in Sec.\ \ref{calcs} one might assume that, 
apart from higher-order crossover effects, both quantities should behave in a 
step-function fashion. On the contrary, we see that they hover around their 
zero-field values for a significant range of $H$, which coincides
with that where our extrapolation gives a straight line. Further on along the
$H$- axis, convergence begins to deteriorate. Taking into account $(i)$ the
conclusion that straight-line sections of PRG curves are finite-size
effects; $(ii)$ the inescapable distortion imposed by them onto
our constant-$H$ extrapolations; plus $(iii)$ the fact that the true critical
line is only expected to be straight {\em at} the origin, where it is joined
by the KT line, we tentatively interpret the plateau-like behaviour of $c$ and
$\eta$ as a manifestation of the KT phase in an artificial, finite-size-induced
fashion. We have not yet managed to propose a numerical test of this idea; however, as shown in the next Section, a measure of self-consistency of the
argument is found in SPAF-3, where both the KT phase and the anomalous
behaviour of $c$ and $\eta$ are absent. Note also that, on general grounds,
estimates of $L/\pi\xi(T,H)$ at $T < T_C^{extr}(H)$ (as $T_c^{real}$ must be)
would certainly
produce numerical values {\em smaller} than those displayed as $\eta$ in
the Figure, which is not inconsistent with the expected $\eta =4/15$. Finally,
as regards $0.10 \leq H < 1.0$ (at which upper extremity one already has
the results of Ref.~\onlinecite{Noh92}), the above-mentioned difficulties with
$L \to \infty$ extrapolations of PRG curves translate into unsurmountable
obstacles to estimations of $c$ and $\eta$.

\section{Square Lattice 3-state Potts Antiferromagnet}
\label{SQ}
We begin the discussion of SPAF-3 by examining the point $T=H=0$, where
strips of maximum width $L=14$ sites were used. We have
calculated $c$ and $\eta$ and found, after extrapolation,
\begin{equation}
c= 0.999 \pm 0.001 \ \ , \eta = 0.333 \pm 0.001\ \ ,   (T=H=0).
\label{sqh0}
\end{equation}
Owing to the unusually slow convergence of finite-$L$ data for $c$ in this case,
we formed three-point fits with the sets $\{f_l(T_c)\}$, $l= L$, $L-2$, $L-4$.
The sequences
of $c_L$, each estimate resulting from a three-point fit, 
were then extrapolated by a Bulirsch-Stoer algorithm~\cite{bst,malte}, which
essentially amounts to assuming a single-power correction to scaling. 
Our best fit corresponded to that power being around 2.
 
The unitary value of $c$ has been predicted for the case~\cite{saleur91,shrock}
 and is consistent with continuously-varying critical
exponents~\cite{cardy87}, thus one might expect {\em e.g.} a KT phase
on the  $T=0$ axis, by analogy with TIAF. We shall return to this point below.
Our result for $\eta$ apparently contradicts the direct evaluation
of correlation functions of Ref. \onlinecite{Bakaev92}, which yields
$\eta = 1.33 \pm 0.02$. To explain this, we recall the prediction of
Ref. \onlinecite{dN82} which, for SPAF-3 with only first-neighbour
interactions (in their language: $v=1$, $\mu=2\pi/3$, $y_K =1/2$), reads:
\begin{equation}
G(r) \simeq {A \over r^{4/3}} \pm {B \over r^{1/3}}
\label{dN}
\end{equation}
\noindent where $G(r)$ is the critical ($T=0$) spin-spin correlation
function at distance $r$; the sign of the second term depends on whether
the two sites are on the same sublattice~\cite{dN82}, that is,
it is associated to the staggered magnetization. 
The result of Ref. \onlinecite{Bakaev92} is for correlations between spins
on opposite corners of $N \times N$ finite lattices with $N \leq 15$,
thus it gives the decay of the uniform magnetization, dominated at
short distances  by the first term of Eq.
(\ref{dN}). Indeed, a plot of their data in the form $r^{4/3} G(r)$ against
$r$ appears to approach a straight line for large $r$,
 with $A \simeq 1.4$ and $B \simeq 5 \times 10^{-3}$.
On the other hand, by relying on the amplitude-exponent
relation given by conformal invariance~\cite{cardy84}, our approach
automatically picks up the behaviour of the smallest gap (or longest
correlation length) of the transfer matrix, which indeed couples to the
staggered magnetization. These 
considerations were very recently rederived via a height representation of the
model~\cite{burton} and confirmed by numerical work~\cite{salas}.
In the present
case, our estimate is entirely consistent with the second term of 
Eq. (\ref{dN}), and also with Monte Carlo work~\cite{salas,Wang90}. 

We have paid special attention to the shape of the critical curve near the
origin. Throughout the range $0.002 \leq H \leq 0.011$, we managed good fits to
Eq.~(\ref{deltatc}) with $\psi$ in the range $4.9 - 6.7$. For now, we 
concentrate on the analysis of that region.

We recall that, although it is agreed that in zero field the system
is critical only at $T=0$, there seems to be no consensus about how the
correlation length diverges, except in that an exponential singularity
$\xi (T \to 0, H=0) \simeq \exp (a/T^x)$ is present. The value of $x$ has 
been variously estimated as 1.3 (by analysis of the Roomany-Wyld
approximant~\cite{rw} in a transfer-matrix calculation~\cite{Night82}, 
and Monte Carlo (MC) work~\cite{Wang90}); 1 (further Monte-Carlo 
work~\cite{Ferreira95}) and
3/4 (conformal invariance arguments coupled with an analysis of the 
eigenvalue spectrum of the transfer matrix~\cite{saleur91}).

If we assume that, along the $T=0$ axis, $H$ is a relevant variable with
scaling index $y_H$, a standard crossover argument~\cite{Kinzel81} implies
that on the critical curve $T_c \sim |\ln H |^{-1/x}$. If, on the
other hand, an extended critical phase is present as in TIAF, the results
of Section \ref{TRI} indicate that such shape is unlikely to be
found. To be fair, we must point out that there is no compelling symmetry-based
argument (such as vortex unpinning for TIAF~\cite{Blote91,Blote93})
which leads one to infer the possible existence of a soft phase here.

\begin{figure}
\epsfxsize=8,5cm
\begin{center}
\leavevmode
\epsffile{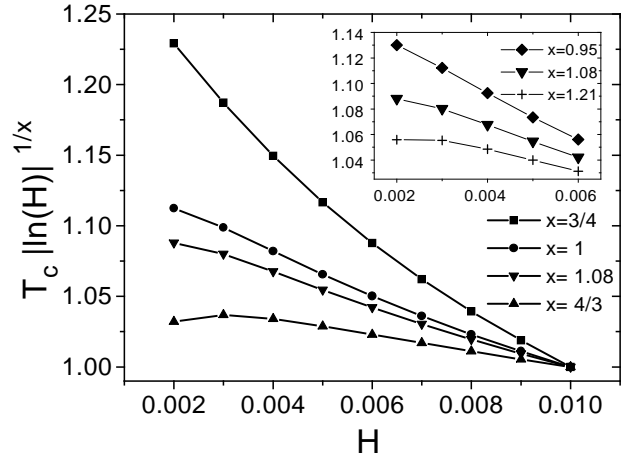}
\caption{Plots of $T_c |\ln H|^{1/x}$ for SPAF-3 for $x= 3/4, 1, 4/3$
and $1.08$. Insert: behaviour of plots for $x=0.95, 1.08$ and $1.21$
from which our central estimate and respective error bars, $x= 1.08 \pm 0.13$
have been extracted (see text). All curves
normalized to one at $H=0.010$. 
Here $T_c$ stands for ($L \to \infty$) extrapolated values. Error bars
coming from extrapolation are smaller than symbol sizes.}
\label{fig:sqpd} 
\end{center}
\end{figure}

In Figure \ref{fig:sqpd} we show our results for $T_c |\ln H|^{1/x}$
against $H$ for
$x=3/4, 1, 4/3$ as well as our best fit for an asymptotically horizontal line as
$H \to 0$, which corresponds to $x= 1.08 \pm 0.13$. This estimate and
its respective error bar are based
on analysis of the insert of the Figure: the $x = 1.21$ curve flattens at
$H \simeq 0.002$ (the lowest field we can reach), so those for $x > 1.21$
certainly bend downwards before touching the vertical axis.
Analogously, the $x = 0.95$ curve is roughly
straight, so those with $x < 0.95$ will be concave upwards. The central
estimate is taken as the average of these upper and lower limits.

Thus we conclude that $(i)$ there is no numerical evidence of an extended
critical phase at $(T=0, H \neq 0)$; $(ii)$ our data are consistent with an
infinite slope of the critical curve at the origin, meaning (via the
crossover argument above) that $(iii)$ the zero-field correlation length
diverges with $T \to 0$ as $\xi (T \to 0, H=0) \simeq \exp (a/T^x)$,
$x= 1.08 \pm 0.13$. 

Of all previously available estimates, the latter value of $x$ is 
only consistent with the
Monte Carlo results of Ref.~\onlinecite{Ferreira95}. Those authors
mention possible logarithmic corrections, which would give an enhanced
effective exponent. This would also be in line with the fact that our
central estimate is slightly above unity.

Further on along the $H$-axis, for
$0.013 \leq H \leq 0.18$ the PRG curves for $L=8$, 10, 12 crossed each other 
at nearly zero angle. In that region, we simply took straight-line fits of
the three respectively values of $T^{\ast}$ against $L^{-1}$ to obtain 
$T_c^{extr}$.
However, for $H \geq 0.2$ monotonic behaviour returned,
once again allowing use of Eq.~(\ref{deltatc}) with $\psi \sim 2.1 - 5.9$.
Additionally,  along either sections of the extrapolated critical line 
matched one another so well across the gap, that they are
joined by continuous lines

\begin{figure}
\epsfxsize=8,5cm
\begin{center}
\leavevmode
\epsffile{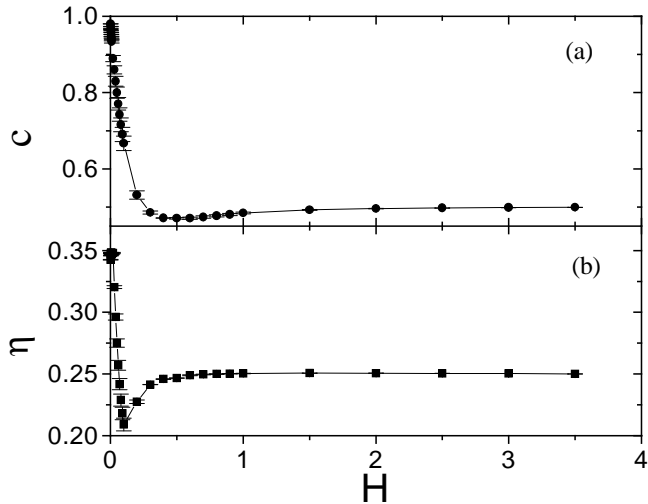}
\caption{Conformal anomaly $c$ (upper curve) and exponent $\eta$ (lower curve)
along the extrapolated critical line of SPAF-3. Expected values
are (a) $H=0$: $c=1$, $\eta=1/3$ (see text); (b) $H \neq 0$: $c=1/2$,
$\eta=1/4$.}
\label{fig:sqetac} 
\end{center}
\end{figure}

In Figure  \ref{fig:sqetac}, estimates of  both  $c$ and $\eta$ along the
extrapolated critical line are displayed. One can see that 
for both quantities, the somewhat
{\em ad hoc} extrapolation procedure in the intermediate-$H$ region produces 
sensible estimates, which join the adjacent sequences
rather smoothly.
 
Further, this time the predictions of Section~\ref{calcs} are seen to hold:
apart from higher-order
crossover effects, both quantities behave close to step-functions,
converging to the respective Ising values  $c=1/2$, $\eta=1/4$.
At $H=0.5$ the curve for $c$ has a minimum. We used strips of width
$L \leq 14$ to produce an accurate estimate both of $T_c$ and $c$,
which turned out as $c=0.47 \pm 0.002$. Thus we conclude that the
(unaccounted for) residual crossover effects produce deviations  of
order at most $\sim 6\%$.

\section{Conclusions}
\label{conc}
We have studied critical properties of field-induced phase transitions of
selected two-dimensional antiferromagnets with discrete-symmetry
order parameters. 
Throughout our work, we attempted to minimise numerical effects
originating from crossover between  different universality classes, by
applying carefully selected procedures both for finite-size calculations
and for extrapolation of finite-size data to the infinite-lattice limit.
For TIAF we did not entirely succeed, owing mainly to residual effects
ascribed to a Kosterlitz-Thouless phase along the
zero-temperature axis. For SPAF-3, where our evidence shows that such phase
is not present, we present results which are clean and unambiguous, for
all quantities investigated.

Our main findings are summarised as follows.
For TIAF, we have shown that the critical line leaves the zero-temperature, zero-field fixed point at a finite angle. Our best estimate
of the slope at the origin is $\left(dT_c/dH\right)_{T=H=0} = 4.74 \pm 0.15$.
For SPAF-3 we provided evidence that the zero-field correlation length
diverges as $\xi(T \to 0, H=0) \simeq \exp (a/T^{x})$, with $x=1.08 \pm 0.13$,
through analysis of the critical curve at $H \neq 0$ plus crossover
arguments. For SPAF-3 we have also ascertained that the conformal anomaly
and decay-of-correlations exponent behave as:   
(a) $H=0$: $c=1$, $\eta=1/3$; (b) $H \neq 0$: $c=1/2$, $\eta=1/4$.

\acknowledgements 
We thank Laborat\'orio Nacional de Computa\c c\~ao Cien\-t\'\i\-fica 
(LNCC) for use of their computational facilities, and
Brazilian agencies  CNPq, FINEP, and FAPERJ for financial support. 
SLAdQ thanks the Department of Theoretical Physics
at Oxford, where parts of this work were carried out, for the hospitality;
the cooperation agreement between Conselho Nacional de Desenvolvimento
Cient\'\i fico e Tecnol\'ogico (CNPq) and  
the Royal Society for funding his visit; and Z. R\'acz for interesting 
discussions.

\begin{table}
\caption{ Slopes $S_L$ and upper limits $H_{max}(L)$ 
of straight-line portions of approximate (PRG) critical lines for TIAF.
From data displayed in Figure \protect{\ref{fig:tiafpd}}. Extr. stands for
extrapolated as $L \to \infty$ (see text). $H_{max}(6)$ is omitted, as
extrapolation only took $L=9 - 18$ into account}
\vskip 0.7cm 
 \halign to \hsize{
\hfil#\quad\hfil&\quad\hfil#\quad\hfil&\quad\hfil#\quad\hfil\cr
$L$ & $S_L$ & $H_{max}(L)$\cr \noalign{\smallskip}
 6  & 1.4979 $\pm$ 0.0001  &  ---  \cr
 9  & 2.0414 $\pm$ 0.0002 &  0.170 $\pm$ 0.010  \cr
 12  & 2.4564 $\pm$ 0.0002 &  0.130 $\pm$ 0.010  \cr
 15  & 2.8078 $\pm$ 0.0002  &  0.105 $\pm$ 0.010  \cr
 18  & 3.1200 $\pm$ 0.0010  &  0.090 $\pm$ 0.010  \cr
 Extr.  & 4.74 $\pm$ 0.15   &  0.01 $\pm$ 0.02  \cr
}
\label{tab:tiaf}
\end{table}

\end{document}